\documentstyle[aps,epsfig,preprint]{revtex}

\newcommand{\be}{\begin{equation}}
\newcommand{\ee}{\end{equation}}

\newcommand{\beet}{\begin{equation*}}
\newcommand{\eeet}{\end{equation*}}
\newcommand{\beaet}{\begin{eqnarray*}}
\newcommand{\eeaet}{\end{eqnarray*}}
\newcommand{\bfig}{\begin{figure}}
\newcommand{\efig}{\end{figure}}
\newcommand{\bc}{\begin{center}}
\newcommand{\ec}{\end{center}}

\newcommand{\szz}{\mbox{$\sigma_{zz}$}}
\newcommand{\srr}{\mbox{$\sigma_{rr}$}}
\newcommand{\szr}{\mbox{$\sigma_{zr}$}}
\newcommand{\srz}{\mbox{$\sigma_{rz}$}}
\newcommand{\scc}{\mbox{$\sigma_{\chi \chi}$}}
\newcommand{\src}{\mbox{$\sigma_{r \chi}$}}

\newcommand{\szc}{\mbox{$\sigma_{z \chi}$}}
 
\begin{document}
\draft

\title{Jamming and Static Stress Transmission in Granular Materials}
\author{M.~E. Cates}
\address{ Dept. of Physics and Astronomy, University of Edinburgh\\
JCMB King's Buildings, Mayfield Road, Edinburgh EH9 3JZ, GB.}
\author{J.~P. Wittmer}
\address{D\'epartment de Physique des Mat\'eriaux,
         Universit\'e Lyon I \& CNRS\\
         69622 Villeurbanne Cedex, France.}
\author{J.-P. Bouchaud, P. Claudin}
\address{Service de Physique de l'Etat Condens\'e,
CEA\\ Ormes des Merisiers, 91191 Gif-sur-Yvette Cedex,
France.}
\maketitle
\begin{abstract}
We have recently developed some simple continuum models of static granular
media which display ``fragile" behaviour: they predict that the
medium is unable to support certain types of infinitesimal load (which we call
``incompatible" loads) without plastic rearrangement. We argue that a fragile
description may be appropriate when the mechanical integrity of the medium
arises adaptively, in response to a load, through an internal jamming
process.
We hypothesize that a network of force chains (or ``granular skeleton")
evolves until it can just support the applied load, at which point it comes to
rest; it then remains intact so long as no incompatible load is applied. Our
fragile models exhibits unusual mechanical responses involving hyperbolic
equations for stress propagation along fixed characteristics through the
material. These characteristics represent force chains; their arrangement
expressly depends on the construction history.  Thus, for example, we predict
a large difference in the stress pattern beneath two conical piles of sand,
one poured from a point source and one created by sieving.
\end{abstract}

\vspace*{1cm}
{\bf Granular materials are microscopically heterogenous. Despite this, it is
natural to search for continuum models that can describe their static and
dynamic behaviour. The problem of granular statics implicitly requires
knowledge of the construction history of the medium. At
the microscopic scale, the construction history determines exactly where every
grain is, and how it has been deformed from its original shape. Given this
information, the microscopic forces follow from the local contact mechanics.
But such a microscopic description of granular materials is, in practical
terms, impossible and unlikely to be a useful guide to
their macroscopic behaviour. For example it is often assumed
that if elasticity governs the local contact mechanics,
the continuum behaviour of the assembly as a whole must be
elastic. This may be unjustified: the physics of the granular
assembly involves additional, strongly nonlinear physics -- namely, whether
each contact is actually present or not. If, as we believe, the contact
network is an adaptive structure that has organized itself to support the
specific load applied during construction, it may obey continuum equations
quite unlike those of conventional elastic or elastoplastic media.}

\section{Introduction}\label{sec:Introduction}

In this paper we consider assemblies of cohesionless rough particles, whose
rigidity is sufficient that individual particle deformations remain always
small. Such assemblies are sometimes argued to be governed by the continuum
mechanics of a Hookean elastic solid (perhaps with a
very high modulus). But this implicitly
assumes that each granular contact is capable equally of supporting tensile as
compressive loads. For a cohesionless medium
this is certainly untrue: cohesionless granular assemblies are therefore not
elastic.\cite{Guyon} 
The question is not one of principle, but of degree -- how important
is the prohibition of tensile forces?  This is not completely clear; some
would argue \cite{deGEv} that it represents a negligible effect and that an
elastic model remains basically sound, so long as the mean stresses in the
material remain compressive everywhere. However, a fully elastic
granular assembly would be one in which grains were, effectively, {\em glued}
permanently to their neighbours on first contact. Because the packing is
microscopically disordered, it is possible that, during subsequent loading {\em a
significant fraction} of such contacts would become tensile, even if the load
being applied remains everywhere compressive {\em on average}. If so, the
absence of tensile forces is a major, even dominant, factor.

Notice that the absence of tensile forces is a distinct physical effect
from the one addressed by most elastoplastic continuum theories of
granular media (see, e.g., Ref.~\cite{Savage98}). These are like elastic
models, but they allow for the fact that the ratio of shear and normal
forces at a contact cannot exceed a fixed value determined by a coefficient
of static friction; this is usually assumed to translate to a
similar condition on the bulk stress components acting across any plane. The
resulting plasticity is similar to that arising in metals, for example, and
not related to the prohibition of tensile forces: it applies
equally for cohesive contacts. Of course, in applying such theories to
cohesionless media one should assume that the mean stresses are everywhere
compressive: however, as emphasized above, this constraint {\em does not}
ensure that individual contact forces are {\em all} compressive as is actually
required.

\begin{figure}[hb]
\centerline{\epsfig{file=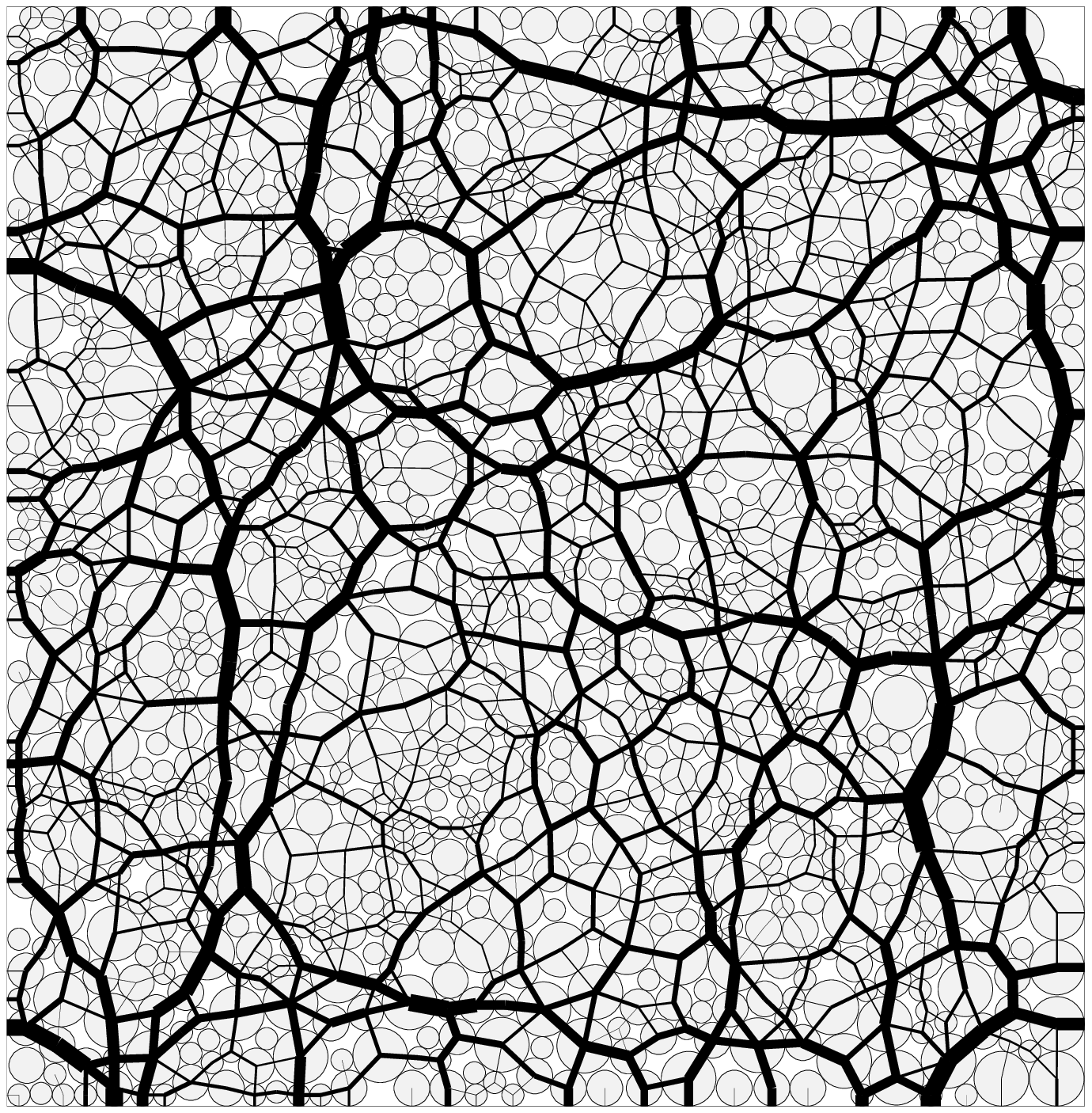,width=80mm,height=80mm,angle=0}}
\vspace*{1cm}
\centerline{\epsfig{file=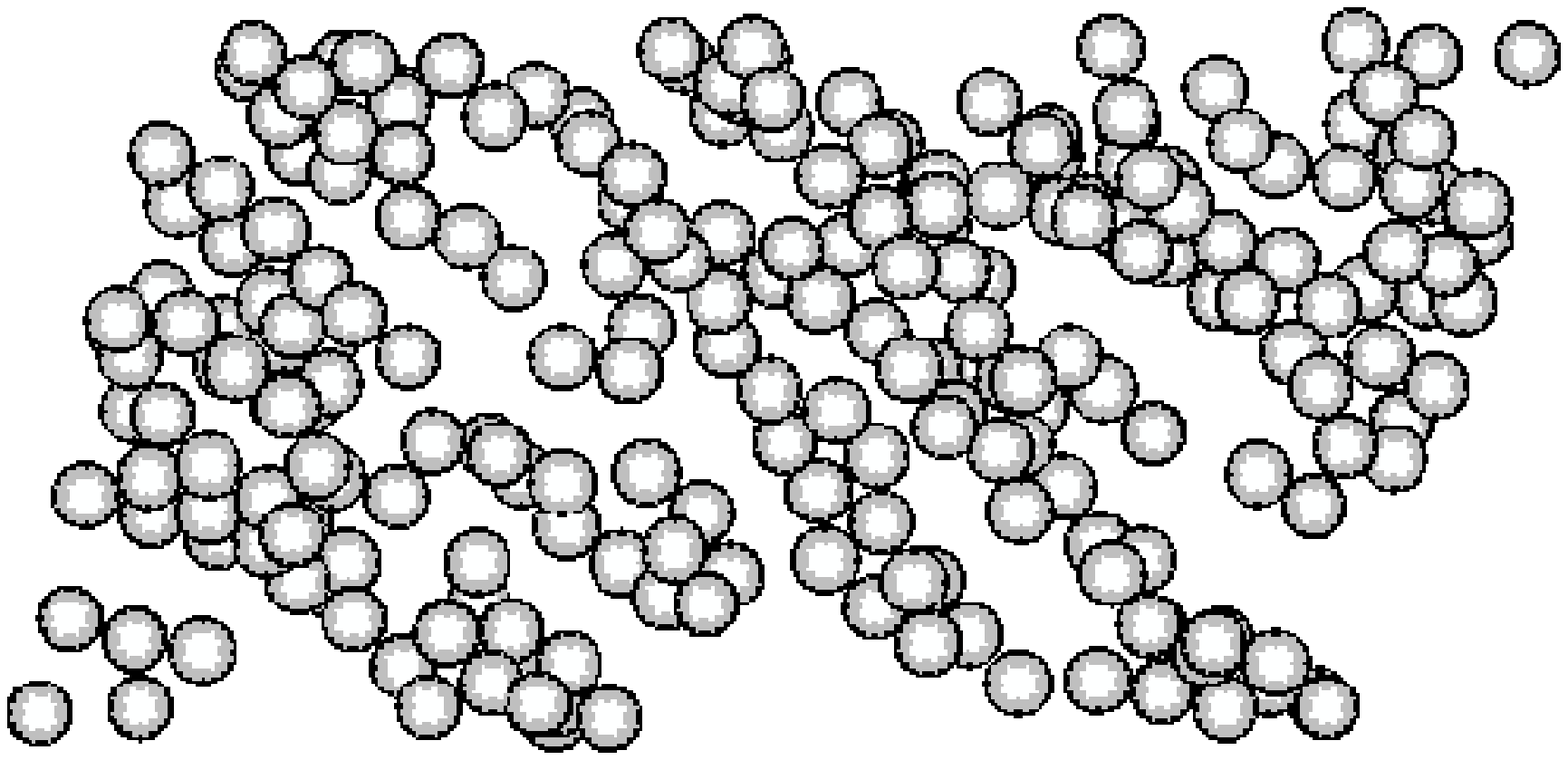,width=80mm,height=50mm,angle=0}}
\vspace*{2cm}
\caption[]{(a) Granular skeleton in two-dimensional 
frictional hard spheres by F.~Radjai {\em et al.}\cite{Radjai} 
(Figure courtesy of F.~Radjai.) 
(b) The jamming transition in a sheared colloid. The data are from a
computer simulation of a hard sphere colloidal suspension
at $\phi = 0.54$ which has been strained to
$\gamma = 0.22$. Shown in the figure are {\em only } those spheres which
have come into very close contact ($\le 10^{-5}$ radius) with
at least one neighbour. As seen from the figure, the contact geometry
is strongly anisotropic and suggests the formation of ``force chains''
running top left to bottom right. 
(The simulation is by J. Melrose, Cavendish Laboratory;
the figure is courtesy of him.) 
\label{fig:snaps}}
\end{figure}

These considerations suggest a physically very different picture of
granular media, already well developed and respected in the sphere of discrete
modelling.\cite{Dantu,Thornton,Radjai}
In this picture, {\em nonlinear} physics is dominant, and the contact network of 
grains is always liable to reorganize as loads are applied:
it is an ``adaptive structure".\cite{Thornton}
The contact network defines a loadbearing ``granular skeleton" 
shown in Fig.~\ref{fig:snaps}(a):
this is often thought of as a network of ``force chains", or roughly 
linear chains of
strong particle contacts, alongside which the other grains play a relatively
minor role in the transmission of stress.\cite{Thornton,Radjai} 
If these ideas are true, the continuum mechanics of the material has to be thought
about afresh. Since this widely-accepted picture of force-chains implies a
microscopically heterogeneous character of the contact network in the
material, it is not necessarily obvious that a continuum description of it is
possible at all. However, we have in recent years developed continuum models
for granular materials which, we now believe, do capture some of the physics
of force chains, and of their geometrical dependence on the construction
history. This interpretation, which has evolved significantly beyond the
empiricism of our early work,\cite{bcc,nature,jphys} is developed below.

The proposal that granular assemblies under gravity cannot properly be
described by the ideas of conventional elastoplasticity has been
opprobiously dismissed in some quarters: we stand accused of ignoring all
that is `long and widely known' among geotechnical engineers.\cite{Savage97}
However, we are not the first to put forward such a subversive proposal.
Indeed workers such as Trollope \cite{Trollope68} and Harr \cite{Harr77} 
have long ago developed ideas of force transfer rules among discrete particles,
not unrelated to our own approaches, which yield continuum equations quite 
unlike those of elastoplasticity.\cite{cargese,PRE,RS,PRL}  
More recently, dynamical {\em hypoplastic} continuum models have been 
developed \cite{Kolymbas} which, as explained by Gudehus \cite{Gudehus97}
describe an `anelastic behaviour without [the] elastic range, 
flow conditions and flow rules of elastoplasticity'. 
Our own models, though not explicitly dynamic, are
similarly anelastic in a specific manner that we describe as ``fragile".

In Section \ref{sec:FM} below, we describe a generic ``jamming" mechanism for the
construction of a granular skeleton that, we argue, points toward fragile
mechanical behaviour. This scenario is related, but not identical, to several
other current ideas in the literature on granular media.
\cite{Guyon,Radjai,Kolymbas,EandO,balledwards,alexander,soc,Moukarzel,Coppersmith,Witten}
These include the emergence of rigidity  by successive
buckling of force chains \cite{alexander} and the concept of mechanical
percolation.\cite{Kolymbas} In particular there is a strong link between 
fragile media and {\em isostatic} models of granular 
assemblies.\cite{Guyon,jnroux} In an isostatic network, the requirement of
force balance at the nodes is enough to determine all the forces acting, so
these can be calculated without reference to a strain or displacement variable.
Isostatic networks require a mean coordination number with a specific
critical value ($z=2d$ with $d$ the dimension of space). In this sense,
isostatic contact networks are ``exceptional", and may appear remote from real
granular materials.

However, it is increasingly clear \cite{Moukarzel,Witten} that almost all
disordered packings of {\em frictionless} spheres actually approach an
isostatic state in the rigid particle limit. Since friction
is ignored, there is still a missing link between this result and the
physics of real granular media -- a link provided by the concept of force
chains, as we show below (Section \ref{subsec:JamFM}). 
More generally, the idea that the
granular skeleton could engineer itself to maintain an isostatic or fragile
state is closely connected with the concepts of self-organized criticality
({\sc soc}) \cite{soc}  (see also Ref.~\cite{staticavalanche}). 
The concepts provide a generic
mechanism whereby an overdamped dynamical system under external forcing can
come to rest at a marginally stable (critical) state. In the {\sc soc}
scenario, this state is characterized by hierarchical (fractal) correlations
and large noise effects (compare Fig.~\ref{fig:snaps}(a)). 
In this article we ignore these
complications and describe only our minimalist, noise-free models of the
granular skeleton; these represent, in effect, regular arrays of force chains.
The effect  of noise on the resulting continuum equations is of great
interest, but these require a separate discussion, 
which is made elsewhere.\cite{cargese,PRE}

\section{Colloids, Jamming and Fragile Matter}\label{sec:FM}

\subsection{Colloids}\label{subsec:Colloid}

We start by describing a simple model of jamming in
a colloid, sheared between parallel plates.\cite{PRL}  This is the simplest
plausible scenario in which an adaptive skeleton forms in response to
an applied load; we believe it sheds much light on the related problem of dry
granular media as discussed in Section \ref{sec:GM} below.  We will first use
it to illustrate some general ideas on the relationship between jamming
and fragility.

Consider a concentrated colloidal suspension of hard particles, confined
between parallel plates at fixed separation, to which a shear stress is
applied (Fig.~\ref{fig:snaps}(b) and \ref{fig:jammed}~(a)). 
Above a certain threshold of stress, this system exhibits enters a regime of 
strong shear thickening; see, e.g., Ref.~\cite{laun}. 
The effect can be observed in the kitchen, by stirring a
concentrated suspension of corn-starch with a spoon. In fact, computer
simulations suggest that, at least under certain idealized conditions, the
material will jam completely and cease to flow, no matter how long the stress
is maintained.\cite{farrmelrose} 
In these simulations, jamming apparently occurs 
because the particles form ``force chains"\cite{Dantu} along the compressional
direction (Fig.~\ref{fig:snaps}(b)). 
Even for spherical particles the lubrication films cannot prevent direct 
interparticle contacts; once these arise, an array or network of force chains 
can indeed support the shear stress indefinitely. 
(We ignore Brownian motion, here and below, as do the simulations; 
this could cause the jammed state to have finite lifetime.)

\begin{figure}[hb]
\centerline{\epsfig{file=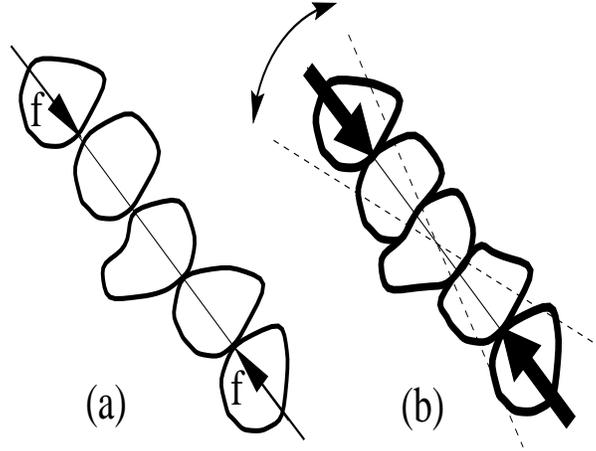,width=80mm,height=60mm,angle=0}}
\vspace*{2cm}
\caption{(a) A force chain of hard particles (any shape) can
statically support only longitudinal compression.
Note that neither friction at the contacts, nor particle aspherity, can
change this ``longitudinal force" rule.
(b) Finite deformability allows small transverse loads
to arise.
\label{fig:paths}}
\end{figure}

To model the jammed state, we start from a simple idealization of a force
chain: a linear string of at least three rigid particles in point contact.
Crucially, this chain can only support loads {\em along its own axis}
(Fig.\ref{fig:paths}~(a)): successive contacts must be
collinear, with the forces along the line of contacts, to
prevent torques on particles within the chain.\cite{EandO}
Note that neither friction at the contacts, nor particle aspherity, can
change this ``longitudinal force" rule. (Particle deformability, however, does
matter; see Section \ref{subsec:Anisotropic} below.)

As a minimal model of the jammed colloid, we therefore take an assembly of such
force chains, characterized by a unique director (a headless unit vector)
$\bf n$, in a sea of ``spectator" particles, and incompressible solvent.
This is obviously oversimplified, for we ignore completely any interactions
between chains, the deflections caused by weak interactions with the spectator
particles, and the fact that there must be some spread in the orientation of
the force chains themselves.
Nonetheless, with these assumptions, in static equilibrium and with
no body forces acting, the pressure tensor $p_{ij}$ (defined as $p_{ij}
=-\sigma_{ij}$, with $\sigma_{ij}$ the usual stress tensor) must obey
\begin{equation}
p_{ij} = P\delta_{ij} + \Lambda\, n_i n_j
\label{eq:nn}
\end{equation}
Here $P$ is an isotropic fluid pressure, and $\Lambda$ ($>0$) a
compressive stress carried by the force chains.

\subsection{Jamming and Fragile Matter}\label{subsec:JamFM}

Eq.~(\ref{eq:nn}) defines a material that is mechanically very unusual. It
permits static equilibrium only so long as the applied compression is along
$\bf n$; while this remains true, incremental loads (an increase or decrease in
stresses at fixed compression axis of the stress tensor) can be
accommodated reversibly, by what is (at the particle contact scale) an elastic
mechanism. But the material is certainly not an ordinary elastic body, for if
instead one tries to shear the sample in a slightly different direction
(causing a rotation of the principal stress axes) static equilibrium cannot be
maintained without changing the director $\bf n$.  Now, $\bf n$ describes the
orientation of a set of force chains that pick their ways through a dense sea
of spectator particles. Accordingly $\bf n$ cannot simply rotate; instead, the
existing force chains must be abandoned and new ones created with a slightly
different orientation. This entails
dissipative, plastic, reorganization, as the particles start to move but then
re-jam in a configuration that can support the new load. The entire contact
network has to reconstruct itself to adapt to the new load conditions; within
the model, this is true even if the compression direction is rotated only by
an infinitesimal amount.

Our model jammed colloid is thus an idealized example of ``fragile
matter": it can statically support applied
shear stresses (within some range), but only by virtue of a
self-organized internal structure, whose mechanical properties have
evolved directly in response to the load itself. Its incremental response can
be elastic only to {\em compatible} loads;
{\em incompatible} loads (in this
case, those of a different compression axis), even if small, will
cause finite, plastic reorganizations. The inability
to elastically support {\em some} infinitesimal loads is 
our chosen technical definition of the term ``fragile".\cite{PRL}

\begin{figure}[hb]
\centerline{\epsfig{file=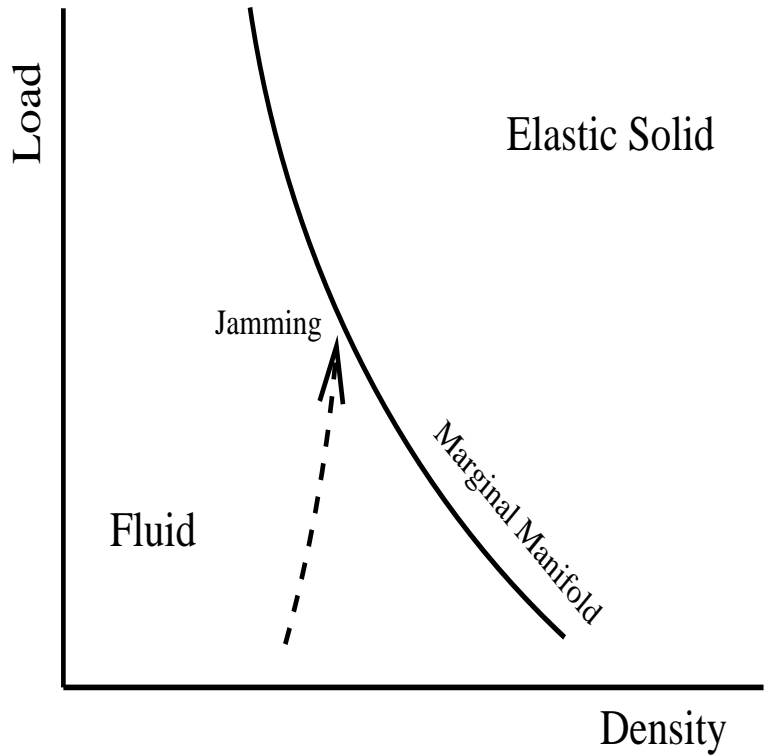,width=100mm,height=100mm,angle=0}}
\vspace*{2cm}
\caption[]{
Schematic ``phase diagram" of a jamming system. 
If, as load or density is increased (dashed arrow), the granular
skeleton arrests on first being able to support the applied load, it can
remain indefinitely on the ``marginal manifold" separating conventional solids
from liquidlike packings. Incompatible loads will move the system to another
point on the same manifold.
\label{fig:phasedia}}
\end{figure}

We argue that jamming may lead {\em generically} to mechanical fragility, 
at least in systems with overdamped internal dynamics. Such a system is likely 
to arrests as soon as it can support the external load; since the load is only 
just supported, one expects the state to be only marginally stable. 
Any incompatible perturbations then force rearrangement; this will leave the system 
in a newly jammed but (by the same argument) equally fragile state. 
This scenario is related, but not identical, to several other ideas in the literature.
\cite{Kolymbas,EandO,balledwards,alexander,soc,Moukarzel} 
The link between jamming and fragility is schematically illustrated in 
Fig.~\ref{fig:phasedia}.

\begin{figure}[hb]
\centerline{
\epsfig{file=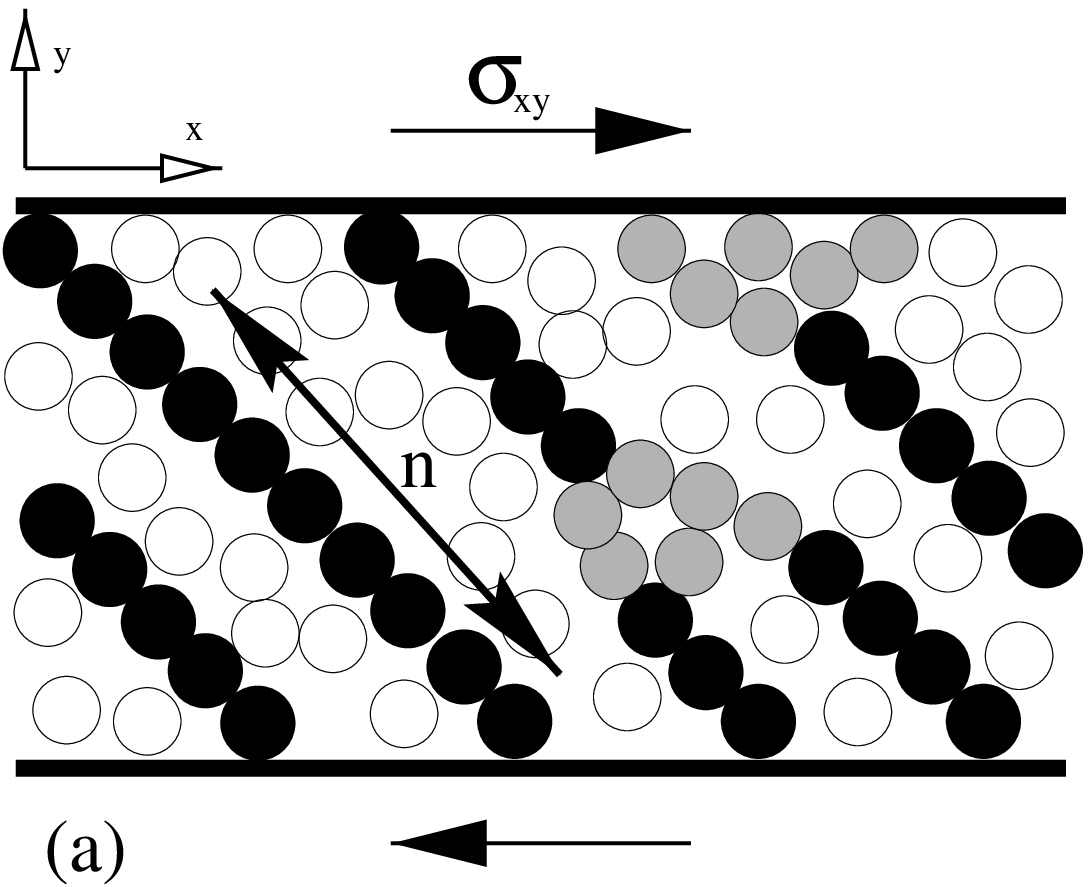,width=80mm,height=65mm,angle=0}
\epsfig{file=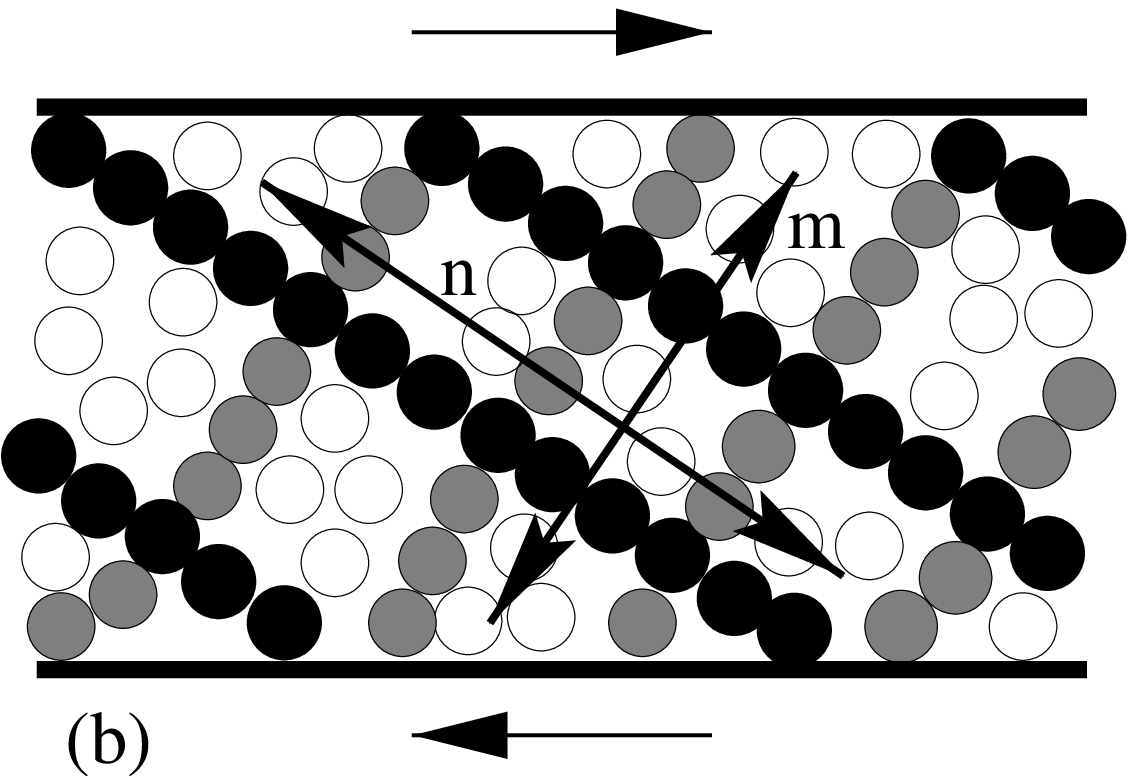,width=80mm,height=55mm,angle=0}}
\vspace*{2cm}
\caption{
(a) A jammed colloid (schematic).
Black: force chains; grey: other force-bearing particles; white:
spectators. (b) Idealized rectangular network of force chains.
\label{fig:jammed}}
\end{figure}

Now consider again the idealized jammed colloid of (Fig.~\ref{fig:jammed}~(a)).
So far we allowed for an external stress field (imposed a the plates) but no
body forces.  What body forces can the system support {\em without} plastic
rotation of the director? Various models are possible.  One is to assume that
Eq.~(\ref{eq:nn}) continues to apply, with
$P({\bf r})$ and $\Lambda({\bf r})$ now varying in space.
If $P$ is a simple fluid pressure, a localized body force can be supported
only if it acts along $\bf n$.
Thus (as in a bulk fluid) no static Green function
exists for a general body force.
(Note that, since Eq.~(\ref{eq:nn}) is already written
as a continuum equation, such a Green function would describes the response to
a load that is localized in space but nonetheless acts on many particles in
some mesoscopic neighbourhood.)
For example, if the particles in Fig.~\ref{fig:jammed}~(a) were to
become subject to a gravitational force along $y$, then the existing force
chains could not sustain this but would reorganize. Applying the longitudinal
force rule, the new shape is easily found to be a catenary, as realized by
Hooke,\cite{Hooke} and emphasized by Edwards.\cite{EandO}  On the
other hand,  a general body force can be supported, in three dimensions,
if there are several different orientations of force chain, possibly
forming a network or ``granular skeleton".\cite{Dantu,Thornton,Radjai,Kolymbas}
A minimal model for this is:
\begin{equation}
p_{ij} = \Lambda_1\, n_i n_j + \Lambda_2\, m_i m_j + \Lambda_3\, l_i l_j
\label{eq:osl1}
\end{equation}
with ${\bf n},{\bf m},{\bf l}$ directors along three
nonparallel populations of force chains; the $\Lambda$'s are
compressive pressures acting along these. Body forces cause
$\Lambda_{1,2,3}$ to vary in space.

We can thus distinguish two levels of fragility, according to whether
incompatible loads include localized body forces ({\em bulk} fragility,
{\em e.g.} Eq.~(\ref{eq:nn})), or are limited to forces acting at the boundary
({\em boundary} fragility, {\em e.g.} Eq.~(\ref{eq:osl1})). In disordered
systems one should also distinguish between macro-fragile responses
involving changes in the {\em mean} orientation of force chains, and the
micro-fragile responses of individual contacts.
%
We expect micro-fragility in
granular materials (see Ref.~\cite{staticavalanche}), although the models
discussed here, which exclude randomness, are only macro-fragile;
in practice the distinction may become blurred. In any case, these
various types of fragility should not be associated too strongly with minimal
models such as Eqs.~(\ref{eq:nn},\ref{eq:osl1}). It is clear that many granular
skeletons
have a complex network structure where many more than three directions of
force chains exist. Such a network may nonetheless be fragile.

Fragility in fact requires any connected
granular skeleton of force chains, obeying the longitudinal force rule ({\sc lfr}), 
to have a mean coordination number $z=2d$ with
$d$ dimension of space (e.g. Fig.~\ref{fig:jammed}~(b) in two dimensions).
This coordination number describes the skeleton, rather than
the medium as a whole; but otherwise it is the same rule as applies for
packings of {\em frictionless} hard spheres. These also obey the {\sc lfr} -- 
not because of force chains, but because there is no
friction. Regular packings of frictionless spheres (which show
isostatic mechanics) have been studied in detail 
recently;\cite{Guyon,cargese,balledwards} and Moukarzel has argued that
disordered frictionless packings of  hard spheres are also generically fragile
\cite{Moukarzel} (see also Ref.~\cite{Witten}). 
These arguments appear to depend only on the {\sc lfr} and the
absence of tensile forces, so they should, if correct, equally apply to any
granular skeleton that is made of force chains of three or more completely
rigid particles.

\subsection{Fixed Principal Axis Model}\label{subsec:FPA}

Returning to the simple model of Eq.~(\ref{eq:osl1}), the chosen values of the
three directors (two in $d=2$) clearly should depend on how the system came to
be jammed (its ``construction history"). If it
jammed in response to a constant external stress,
switched on suddenly at some earlier time, one
can argue that the history is specified purely by
the stress tensor itself. In this case, if one
director points along the major compression axis
then by symmetry any others should lie at
rightangles to it (Fig.~\ref{fig:jammed}~(b)).
Applying a similar argument to the intermediate
axis leads to the ansatz that all three directors
lie along principal stress axes; this is perhaps
the simplest model in $d=3$. One version of this
argument links force chains with the fabric tensor,\cite{Kolymbas} 
which is then taken coaxial with the stress.\cite{Radjai}
(The fabric tensor is the second moment of the orientational distribution
function for contacts and/or interparticle forces.)

With the ansatz of perpendicular directors as just described,
Eq.~(\ref{eq:osl1}) becomes a ``fixed principle axes" ({\sc fpa}) 
model.\cite{nature,jphys,cargese} 
Although grossly oversimplified, this leads to
nontrivial predictions for the jammed state in the colloidal problem, such as
a constant ratio of the shear and normal stresses when these are varied in the
jammed regime. Such constancy is indeed reported by Laun\cite{laun} in
the regime of strong shear thickening; see Ref.~\cite{PRL}.

\section{Granular Materials}\label{sec:GM}

We believe that these simple ideas on jamming and fragility in colloids are
equally relevant to cohesionless, dry granular media constructed from hard
frictional particles. For although the formation of dry granular aggregates
under gravity is not normally described in terms of jamming, it is a closely
related process. 
Indeed, the filling of silos and the motion of a piston in a
cylinder of grains both exhibit jamming and stick-slip phenomena associated
with force chains; see Ref.~\cite{samchains}. And, just as in a jammed colloid, the
mechanical integrity of a sandpile entirely disappears as soon as the load (in
this case gravity) is removed.

\begin{figure}[hb]
\centerline{\epsfig{file=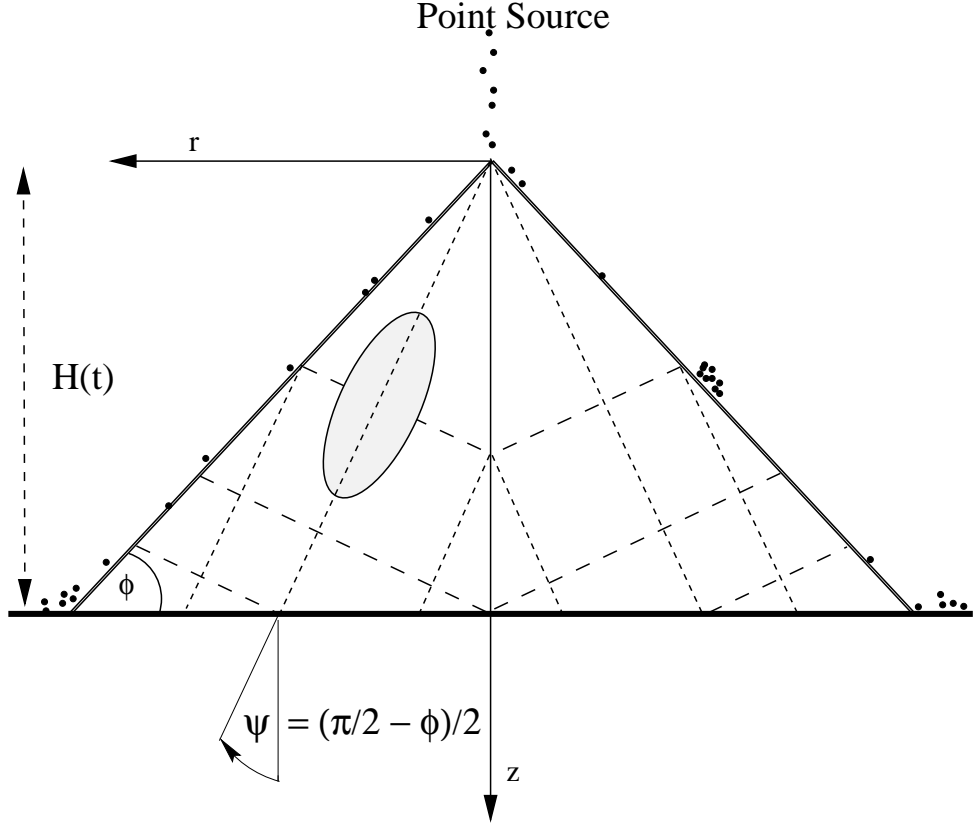,width=130mm,height=110mm,angle=0}}
\vspace*{2cm}
\caption{
Definition of normal construction history of a pile.
The grains fall down from the point source on the pile and roll down the slopes,
which are at the repose angle $\phi$. The height of this pile is $H(t)$.
Sketch of the geometry of the {\sc FPA} model:
The stress ellipsoid has fixed inclination angle $\Psi$; its ellipticity varies
from zero at the centre of the pile to a maximum in the outer region.
The outward and inward stress propagation characteristics are indicated by
short-dashed and long-dashed lines; these are at rightangles and coincident
with the principal axed of the stress ellipsoid.
\label{fig:pilegreen}}
\end{figure}

In the granular context, a model like Eq.~(\ref{eq:osl1}) can be interpreted by
saying that a fragile granular skeleton of force chains is laid down at the
time when particles are first buried at a free surface (see Fig.~\ref{fig:pilegreen}); 
so long as subsequent
loadings are compatible with this structure, the skeleton will remain intact
-- if not grain by grain, then at least in its average properties. If in
addition the skeleton is rectilinear (perpendicular directors) this forces the
principal axes to maintain forever the orientation they had close to the free
surface ({\sc fpa} model). However, we do not insist on this last property and
other models, which correspond to an oblique family of directors in
Eq.~(\ref{eq:osl1}), will be described below.\cite{jphys,PRL,RS}

In what follows we review in more detail the nature of our fragile models and
the role they might play within a continuum mechanical description of granular
media. We will mainly be concerned with the {\em standard sandpile}, which we
define to be a conical pile, constructed by pouring cohesionless hard grains
from a point source onto a perfectly rough, rigid support 
as shown in Fig.~\ref{fig:pilegreen}.
We assume that this
construction leads to a series of shallow surface avalanches whereby all
grains have come to rest, at the point of burial, very close to the free
surface of the pile. (Very different conditions may apply for {\em wedges} of
sand; see Ref.~\cite{RS}.) An alternative history is the {\em sieved pile},
which is a cone created by sieving a series of concentric discs one on top of the
other. In the standard sandpile, it is well known that the vertical normal
stress has a minimum, not a maximum, beneath the apex.\cite{SN,Huntley}
A striking feature of our modelling approach is that it
not only accounts for this ``stress-dip" reasonably well, but predicts that it
should be entirely absent for a sieved pile. This proposal is currently being
subject to careful experimental verification.\cite{Vanel99}

\subsection{Continuum Modelling of Granular Media}\label{subsec:ContMod}

The equations of stress
continuity express the fact that, in static equilibrium, the
forces acting on a small element of material must balance. For a
conical pile of sand we have, in
$d=3$  dimensions,
\be \partial_r\srr + \partial_z\szr = \beta {(\scc-\srr)/r}\label{eq:cont1}\ee
\be\partial_r\srz + \partial_z\szz = g - \beta\srz/r\label{eq:cont2}\ee
\be\partial_\chi\sigma_{ij} = 0\label{eq:cont3}\ee where $\beta = 1$.
Here
$z,r$ and $\chi$ are cylindrical polar coordinates, with $z$ the
downward vertical. We take $r=0$ as a symmetry axis, so that
$\src=\szc=0$; $g$ is the force of gravity per unit volume;
$\sigma_{ij}$ is the usual stress tensor which is symmetric in
$i,j$. The equations for
$d=2$ are obtained by setting $\beta = 0$ in (\ref{eq:cont1},\ref{eq:cont2})
and suppressing (\ref{eq:cont3}). These describe a wedge of constant
cross section and infinite extent in the third dimension.

The Coulomb law states that, at any point in a cohesionless
granular medium, the shear force acting across any plane must be
smaller in magnitude than $\tan\phi$ times the compressive normal force.
Here $\phi$ is the angle of friction, a material parameter which,
in simple models, is equal to the angle of repose.
We accept this here, while noting that
(i) $\phi$ in principle depends on the texture (or fabric) of the
medium and hence on its construction history;
(ii) for a highly anisotropic packing, the existence of an
orientation-independent $\phi$ is questionable;
(iii) the identification of $\phi$ with the repose angle ignores some
complications such as the Bagnold hysteresis effect
(which may in turn be coupled to density changes).
Setting these to one side, we note that the Coulomb law is an {\em inequality}:
therefore, when combined with stress continuity, it cannot lead to closed
equations for the granular stresses. To close the system of equations, further
assumptions are clearly required. One choice is to assume that the material is
an elastic continuum wherever it does not violate the Coulomb condition. (This
is the simplest possible type of elastoplastic model.) A second choice it to
treat the Coulomb condition as though it were an {\em equality}. This is the
basis of the so-called ``rigid plastic" approach to granular media. We return
to both of these modelling schemes after first describing our own approach.

\subsubsection{Constitutive Relations Among Stresses}\label{subsec:Constitutive}

We view cohesionless granular matter as assembly of rigid particles held up
by friction. The
static indeterminacy of frictional forces can, we argue, then be
circumvented by assuming the existence of some local {\em constitutive relations} 
(c.r.'s) among components of the stress tensor.\cite{bcc,nature,jphys}
The c.r.'s among stress components are taken to
encode the network of contacts in the granular packing geometry; they therefore
depend explicitly on its
construction history. The task is then to postulate and/or justify physically
suitable c.r.'s among stresses, of which only one (the {\em primary} c.r.)
is required for systems with two dimensional symmetry, such as a
wedge of sand;  for a three dimensional symmetric assembly
(the conical sandpile) a secondary c.r. is also needed.

The above nomenclature has caused confusion to some commentators on our work.
In solid mechanics the term `constitutive relation' normally refers to a
material-dependent equation relating stress and strain. In fluid mechanics one
has instead equations relating stress and (in the general case of a
viscoelastic fluid) strain-rate history. Instead, our models of granular media
entail equations relating stress components to one another, in a manner that
we take to depend on the construction history of the material. Clearly such
equations are intended to describe constitutive properties of the medium: they
relate its state of stress to other discernable features of its physical
state. We see no alternative to the term `constitutive relations' for such
equations.

In the simplest case, which is the {\sc fpa} model\cite{nature,jphys}
one hypothesizes that, in each material
element, the orientation of the stress ellipsoid became `locked' into
the material texture at the time when it last came to rest, and does not
change in subsequent compatible loadings.
This is a bold, simplifying assumption, and it may
indeed be far too simple, but it exemplifies the idea of having a local
rule for stress propagation that depends explicitly on construction
history. At first sight the idea of `locking in' the principal axes seems to
contradict the conception of an adaptive granular skeleton which can rearrange
in response to small incremental loads. Remember though that this `locking in' 
only applies for compatible loads -- those which the existing skeleton can support. 
Any incompatible load will cause reorganization. We therefore require that
any incompatible loads are specified in defining the construction history of
the material.

For the standard sandpile geometry (see Fig.~\ref{fig:pilegreen}), 
where the material comes to
rest on a surface slip plane, such loads do not in fact arise after material
is buried. The {\sc fpa} constitutive hypothesis then leads to the following
primary c.r. among stresses:
\be\srr = \szz -2\tan\phi \,\,\szr \label{eq:fpa1}\ee 
where $\phi$ is the angle of repose.
Eq.(\ref{eq:fpa1}) is algebraically specific to the case
of a standard sandpile created from a point source by a series of avalanches
along the free surface.  The conceptual basis of the {\sc fpa} model is not so
narrow: indeed, we applied it already to jammed colloids in Section \ref{sec:FM}. 
More generally the {\sc fpa} model is arguably the simplest
possible choice for a history-dependent c.r. among stresses; but this does not
mean it will be sensible in all geometries.

A consequence of Eq.~(\ref{eq:fpa1})  for a standard sandpile, is that the
major principal axis everywhere bisects the angle between the
vertical and the free surface. It should be noted that in
cartesian coordinates, the {\sc fpa} model reads:
\be\sigma_{xx} = \szz -2\,\hbox{\rm sign}(x)\tan\phi
\,\,\sigma_{xz}\label{eq:fpa2}\ee 
where $x=\pm r$ is horizontal. From Eq.~(\ref{eq:fpa2}), the {\sc fpa}
constitutive relation is seen to be discontinuous on the central axis of
the pile: the local texture of the packing has a singularity on the
central axis which is reflected in the stress propagation rules of the
model. (This is physically admissible since the centreline
separates material which has avalanched to the left from material which
has avalanched to the right.) The paradoxical requirement, on the centreline,
that the principal axes are fixed simultaneously in two
different directions has a simple resolution:
the stress tensor turns out to be isotropic there. See Fig.~\ref{fig:pilegreen}. 
The constitutive singularity leads to an `arching' effect for the standard sandpile, 
as previously put forward by Edwards and Oakeshott\cite{EandO}
and others.\cite{Trollope68,Burman80}

\begin{figure}[hb]
\centerline{\epsfig{file=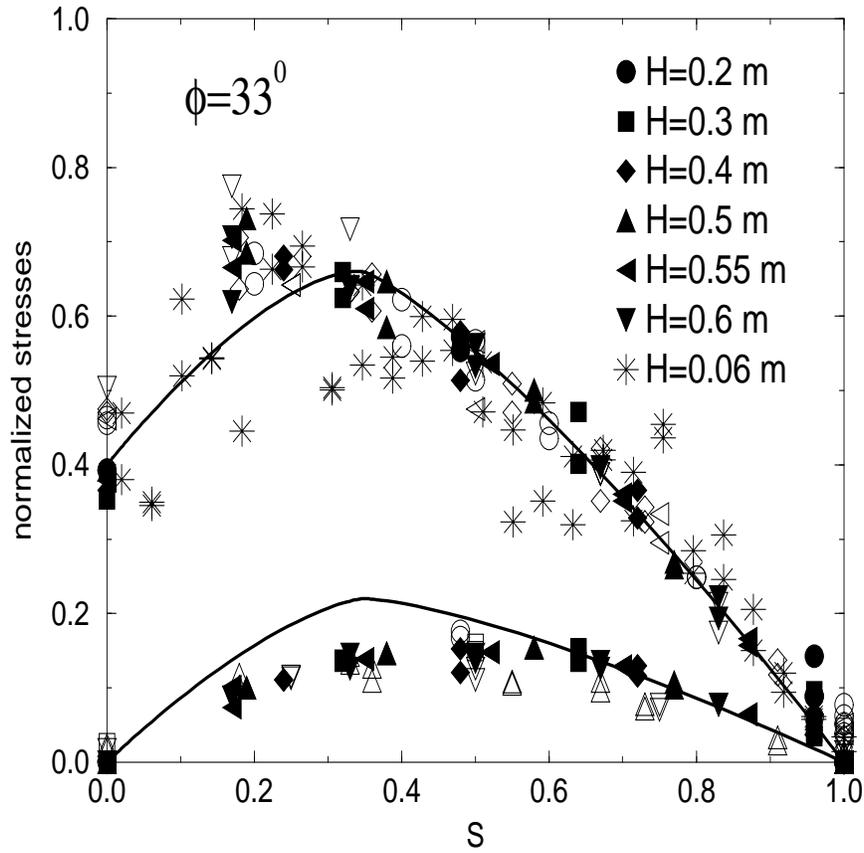,width=130mm,height=130mm,angle=0}}
\vspace*{2cm}
\caption[]{Comparison of {\sc FPA} model
Using a uniaxial secondary closure\cite{nature,jphys} 
with scaled experimental data of Smid \& Novosad\cite{SN} and
\mbox{(*)} that of Brockbank {\em et al.}\cite{Huntley}
which was averaged over three piles.
Upper and lower curves denote normal and shear stresses.
The data is used to calculate the total weight of the pile which is then
used as a scale factor for stresses. The horizontal coordinate
$S=r \tan(\phi)/H$ is scaled by the pile height $H$.\label{fig:dip}}
\end{figure}

The {\sc fpa} model is one of a larger class of {\sc osl}
(for ``oriented stress linearity'') models
in which the primary constitutive relation (in the sandpile
geometry) is, in Cartesians
\be \sigma_{xx} = \eta \szz + \mu\, \hbox{\rm sign}(x)\,\,\sigma_{xz}
\label{eq:osl2}\ee with $\eta,\mu$ constants.
Note that the boundary condition, that
the free surface of a pile at its angle of repose $\phi$ is a slip plane,
yields one equation linking
$\eta$ and $\mu$ to $\phi$; thus, for a sandpile geometry, the {\sc osl}
scheme represents a one-parameter family of primary c.r.'s.
The {\sc osl} models were developed \cite{jphys} to explain experimental
data on the stress distribution beneath a standard sandpile.
\cite{SN,Huntley,Jokati79} 
With a plausible choice of secondary c.r. 
(of which several were tried, with only minor differences resulting), 
the experimental data (Fig.~\ref{fig:dip}) is found to support models in the 
{\sc osl} family with $\eta$ close, but perhaps not exactly equal, to unity 
(the {\sc fpa} value). This is remarkable, in view of the radical simplicity of 
the assumptions made.

As explained by Wittmer {\em et al.},\cite{nature,jphys} the {\sc osl} models, 
combined with stress continuity (Eq.~\ref{eq:osl2}) yield hyperbolic equations 
having {\em fixed characteristic rays} for stress propagation. In fact they are 
wave equations;\cite{bcc,jphys}
moreover they are essentially equivalent to Eq.~(\ref{eq:osl1}), 
with (in general) an oblique triplet of directors
(these become mutually orthogonal only in the case of {\sc fpa}). The
constitutive property that {\sc osl} models describe is that these
characteristic rays (and not, in general, the principal axes) have
orientations that are `locked in' on burial of an element, and do not change
when a further compatible load is applied. As demonstrated already in 
Section \ref{sec:FM}, there is every reason to identify such characteristics, 
in the continuum model, with the mean orientations of force chains in the underlying
material.

Note that unless the {\sc osl} parameter is chosen
so that $\mu = 0$, a constitutive singularity on the central axis, as
mentioned above for the {\sc fpa} case, remains. (The characteristics are
asymmetrically disposed about the vertical axis, and invert discontinuously at
the centreline
$x=0$.) The case $\mu =0$ corresponds to one studied
earlier by Bouchaud {\em et al.},\cite{bcc} and of the {\sc osl} family it is the
only candidate for describing a sieved pile, in which the construction history
clearly cannot lead to a constitutive singularity at the axis of the pile.
The resulting `{\sc bcc}' model could be called a `local Janssen model' in that
it assumes {\em local} proportionality of horizontal and vertical compressive
stresses -- an assumption which, when applied {\em globally} to average
stresses in a silo, was first made by Janssen.\cite{Janssen}
The {\sc bcc} model predicts a smooth maximum, not a dip, in the pressure beneath 
the apex of a pile. 
This is what we expect, therefore, in the case of a sieved pile.\cite{jphys}

\subsubsection{Rigid-Plastic Models}\label{subsec:RPM}

A more traditional, but related, approach is one based on the
(Mohr-Coulomb) rigid-plastic model.\cite{Nedderman} 
To find so-called limit state solutions in this model,
one postulates 
that the Coulomb condition is everywhere obeyed 
{\em with equality}.\cite{Savageclaim}
That is, through every point in the material there passes some plane across 
which the shear force is exactly $\tan\phi$ times the normal force. 
By assuming this, the model allows closure 
(modulo a sign ambiguity discussed below) of the equations for the stress 
without invocation of an elastic strain field.

This limit-state analysis of the rigid plastic model is
equivalent to assuming a `constitutive relation'
(sometimes called `incipient failure everywhere' \cite{jphys}):
\be
\sigma_{rr} = \sigma_{zz}\,{1\over \cos^2\phi}\left[
\sin^2\phi+1 \pm
2\sin\phi\,\,\sqrt{1-(\cot\phi\;\;\sigma_{zr}/\sigma_{zz})^2}\right]
\label{eq:ife}
\ee
whereas the Coulomb inequality requires only that $\sigma_{rr}$
lies between the two values ($\pm$) on the right.
It is a simple exercise to show
that for a sandpile  at its repose angle, only one solution
of the resulting equations  exists in which the sign choice is everywhere
the same. This requires the negative root (conventionally referred to
as an `active' solution) and it shows a hump, not a dip, in the vertical
normal stress beneath the apex. Savage\cite{Savage97}, however, draws attention 
to a `passive' solution, having a pronounced dip beneath the apex.\cite{Savageclaim} 
The passive solution actually contains a pair of matching planes between an inner 
region where the positive root of (\ref{eq:ife}) is taken, and an outer region where 
the negative is chosen. In fact (see Ref.~\cite{RS}) there are more than one such matched
solutions, corresponding to different types of discontinuity in the stress (or
its gradient) at the matching plane and/or the pile centre. Moreover, there is
no physical principle that limits the number of matching surfaces; by adding
extra ones, a very wide variety of results might be achieved.

It is interesting to compare the mathematics, and physics, of Eq.~(\ref{eq:ife})
with that of the {\sc osl} models introduced above. The rigid-plastic model
yields a local c.r. among stresses; like {\sc osl} the resulting equations are
hyperbolic. It also exhibits fragility: because a yield plane passes through
every material point, certain incremental loads will cause reorganization.
Therefore, anyone who defends the rigid-plastic model as a cogent description
of sandpiles cannot reasonably object to these same features in
our own models.  Conversely, we cannot object in principle to
a model in which a  Coulomb yield plane passes through every material point.
However, we still see no reason why it should be a good model;\cite{jphys} 
in particular we cannot see how to make a link between the
characteristic rays in this model (which are always load dependent) and the
underlying geometry of the contact network in the medium. In contrast, this
link arises naturally in the {\sc osl} framework.

\subsubsection{Elastoplastic models}\label{subsec:ElastPlast}

The simplest elastoplastic models assume a material
in which a perfectly elastic behaviour at small strains is conjoined
onto perfect plasticity (the Coulomb condition with equality) at larger
ones. In such an approach to the standard sandpile, an inner elastic region
connects onto an outer plastic one. In the inner elastic region the stresses
obey the Navier equations, which follow from those of Hookean elasticity by
elimination of the strain variables. The
corresponding strain field is usually not discussed, but tacitly treated as
infinitesimal: the high modulus limit is taken. It has been argued that, for a
sandpile on a rigid support {\sc fpa}-like solutions can be found within a
purely elastoplastic model, at least in two dimensions.\cite{Cantelaube}
However, since these show a cusp in the vertical stress on the centreline, they
imply an infinitesimal displacement field incompatible with a continuous
elasticity across the midline.\cite{Savage98} On the other hand,
it is possible to obtain a stress dip, in an elastoplastic model, by assuming
that the supporting base is not rigid but subject to basal sag.\cite{Savage98}
This explanation cannot explain the data of Huntley\cite{Huntley} 
which involves an indentable (rather than sagging) base.\cite{RS}
Moreover, it would predict a similar dip for a sieved pile, unlike or own models; 
experiments on this point are now available, and suggest that indeed 
no dip is seen in this case.~\cite{Vanel99}

Objections to the elastoplastic approach to modelling sandpiles can also be
raised at a much more fundamental level.\cite{PRL,RS} Specifically, to
make unambiguous predictions for the stresses in a sandpile, these models
require boundary information which, at least for the simpler models, can be
given no clear physical meaning or interpretation. We return to this point
below.

\subsection{Fragile vs. Elastoplastic Descriptions}\label{subsec:Boundary}

\subsubsection{Problems definining an elastic strain}\label{sec:Strain}

In the {\sc fpa} model and its relatives, strain variables are not considered.
No elastic modulus enters, and there is no intrinsic stress scale. The
resulting predictions for a conical pile therefore obey what is
usually called radial stress-field ({\sc rsf}) scaling. Formally one has for
the stresses at the base
\be
\sigma_{ij} = gh\, s_{ij}(r/ch) \label{eq:rsf}
\ee
where $h$ is the pile height, $c = \cot\alpha$  and $s_{ij}$ a reduced
stress: $\alpha$ is the angle between the free surface and the horizontal so
that for a pile at repose, $\alpha = \phi$.
This form of {\sc rsf} scaling, 
which involves only the forces at the base,\cite{Evesque97} 
might be called the `weak form' and is
distinct from the `strong form' in which Eq.~(\ref{eq:rsf}) is obeyed
also with $z$ (an arbitrary height from the apex) replacing $h$ (the
overall height of the pile). Our {\sc osl} models obey both forms; only the
weak form has been tested directly by experiment but it is well-confirmed in
many systems (Smid and Novosad\cite{SN}, Huntley\cite{Huntley}).

The observation of {\sc rsf} scaling, to
experimental accuracy, suggests that elastic effects {\em
need not be considered explicitly}.
This does not of itself rule out
elastic or elastoplastic behaviour which, at least in the limit of
large modulus, can also yield equations for the stress from which the bulk
strain fields, and hence also the modulus, cancel. (Note that it is tempting,
but entirely wrong, to assume that a similar cancellation occurs at the
{\em boundaries} of the material; we return to this below.)
The cancellation of bulk strain fields in elastoplastic
models disguises a serious problem in their
application to the standard sandpile and related geometries.\cite{RS,PRL} 
The difficulty is this: there is no obvious definition of
strain or displacement for such a construction history. To define a
physical displacement or strain field, one requires a {\em reference state}. In
(say) a triaxial strain test (see, e.g., Ref.~\cite{Wood90}) an initial state is made by
some reproducible procedure, and strains measured from there. The elastic part
is identifiable in principle, by removing the applied stresses (maintaining an
isotropic pressure) and seeing how much the sample recovers its shape. In
contrast, a pile constructed by pouring grains onto its apex is not made by a
plastic and/or elastic deformation from some initial reference state of the
same continuous medium. The
problem of the missing reference state occurs whenever the solidity of the body
itself arises purely because of the load applied. Thus,
for the jammed colloid considered in Section \ref{sec:FM} above, the
unloaded state is simply a fluid. For the sandpile, it is grains floating
freely in space. On cannot satisfactorily define an elastic strain with respect
to either of these reference states.

A route to defining a strain variable does however exist,\cite{deGEv}
so long as one ignores the fact that tensile forces are prohibited. In effect,
one assumes that when grains of sand arrive at the free surface, each one forms
permanent or``glued" elastic contacts with its neighbours;\cite{PRL}
this contact network can then, by assumption, elastically support arbitrary
incremental loads. This is an admissable physical hypothesis, though
contradictory to our own hypothesis of an adaptive, fragile granular skeleton.
We do not yet know which hypothesis is more correct; the test of this lies
in experiment. (It does not lie in a sociological comparison of how physicists and
engineers approach their work, as offered by Savage.\cite{Savage98})

\begin{figure}[hb]
\centerline{
\epsfig{file=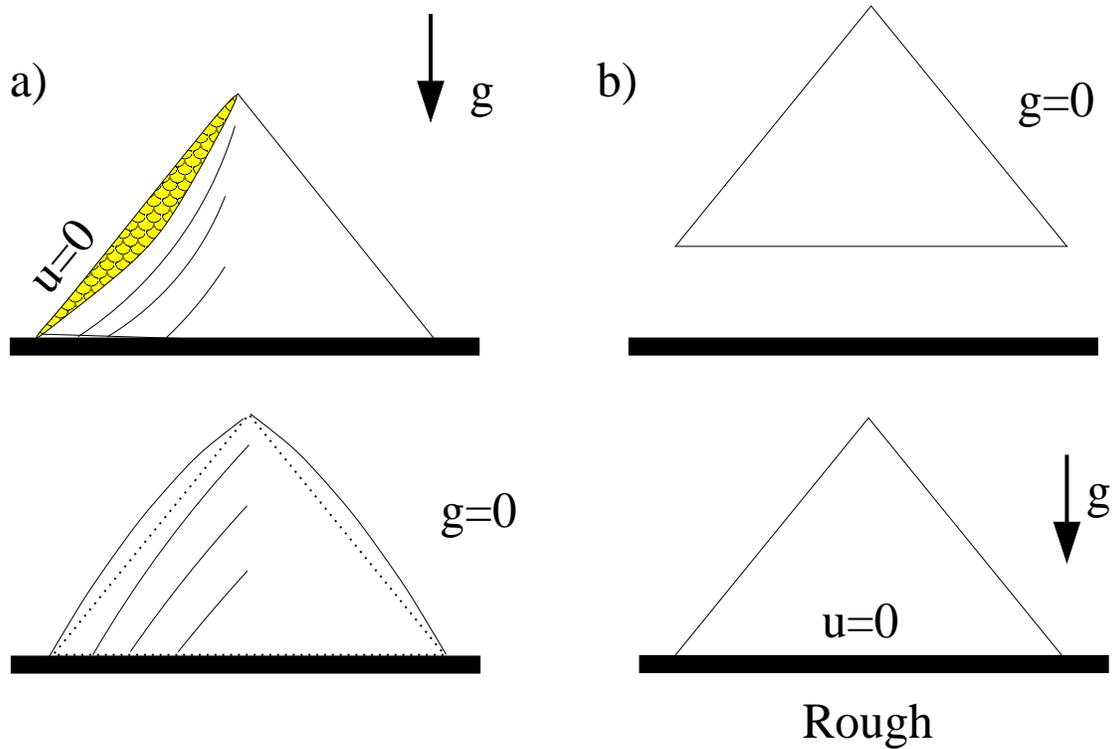,width=150mm,height=100mm,angle=0}}
\vspace*{2.0cm}
\caption{
(a) Quenched stresses in an {\em elastic} sandpile.
Layers are added in a state of zero stress to a pre-existing, 
gravitationally loaded pile.
Such a pile (if gravity is removed) will spring into a new shape,
characterized by a nonzero internal stress field
which includes tensile streses. These require rearrangements if the grains
are cohesionless; the response to gravity is intrinsically nonlinear.
(b) The `floating' model of a sandpile.
An unstrained, isotropic elastoplastic cone is brought into contact with a 
rough surface and gravity then switched on. This is the only construction
history we can think of that completely avoids quenched stresses in the 
formation of the pile.
\label{fig:floating}}
\end{figure}

If the ``glued pile" model is correct, then a strain variable is defined from
the relative displacement that has occurred between adjoining
particles since the moment they first were glued together.\cite{deGEv}
However, the resulting displacement field, found by integrating the strain, is
unlikely to be single-valued. Put differently, if a glued assembly is
created under gravity and then gravity is switched off, it will revert to a
state in which there are residual elastic strains throughout the material,
even though there is now no body force acting (Fig.~\ref{fig:floating}(a)). 
This is because the particle contact network was itself created under 
partially-loaded conditions. 
Many elastic and elastoplastic calculations, such as
all those reviewed by Savage,\cite{Savage98} entirely ignore the problem of 
quenched stresses, and therefore embody an implausible ``floating
model" of a sandpile shown in Fig.~\ref{fig:floating}~(b).

Note that these effects do not become small when the limit of a large
modulus is taken; the quenched stresses remain of order the stress that
was acting during formation, and can take both signs (tensile as well
as compressive).\cite{Moukarzel} So, if one creates a glued pile under gravity
and then slowly switches off the body force, tensile forces will arise long
before $g$ has gone to zero. In this sense, the response to gravity of a 
cohesionless pile is completely nonlinear. Correspondingly, in an unglued pile, 
no smooth deformation can connect the state of a pile created under gravity with an
unloaded state of the same contact network: as the load is removed, such
a pile will undergo large-scale reorganization.

\subsubsection{Boundary conditions and determinacy in hyperbolic models}
\label{subsec:Hyperbolic}

Models, such as {\sc osl}, that assume local
constitutive equations among stresses provide hyperbolic differential
equations for the stress field. Accordingly, if one specifies a
zero-force boundary condition at the free (upper) surface of a
granular aggregate on a rough rigid support, then any perturbation arising from
a small extra body force (in two dimensions, for simplicity) propagates along
two characteristics passing through this point. In the {\sc osl} models
these characteristics are, moreover, straight lines.
Therefore the force at the base can be found simply by
summing contributions from all the body forces as propagated along
two characteristic rays onto the support; the sandpile problem is, within
the modelling approach by
Bouchaud {\em et al.}\cite{bcc} and Wittmer {\em et al.},\cite{nature,jphys}
mathematically well-posed. There is no need to consider any elastic strain
field and the paradoxes concerning its definition in cohesionless poured sand,
discussed above, do not arise.

\begin{figure}[hb]
\centerline{\epsfig{file=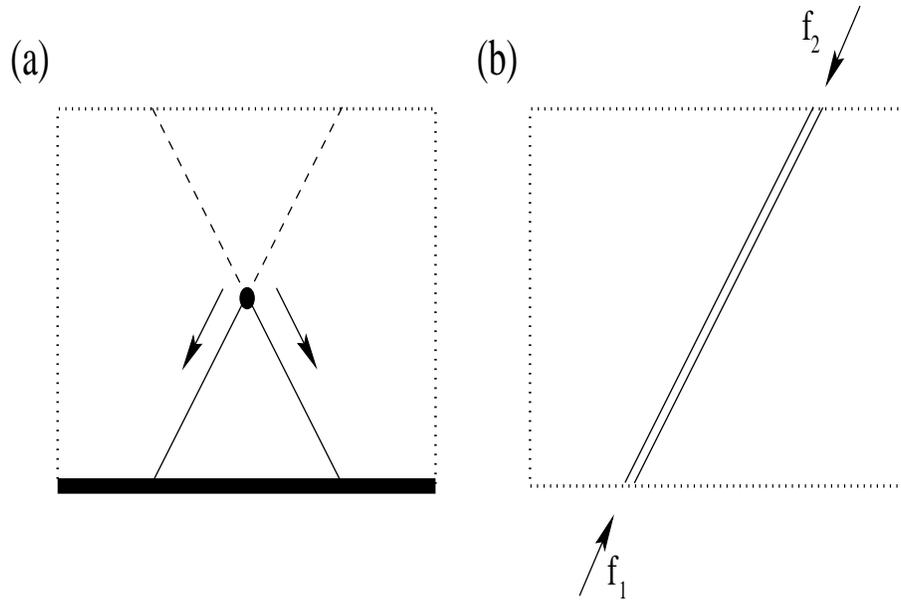,width=120mm,height=80mm,angle=0}}
\vspace*{2cm}
\caption{
(a) The response to a localized force is found by resolving it along
characteristics through the point of application, propagating along those
which do not cut a surface on which the relevant force component is specified.
For a pile under gravity, propagation is only along the downward rays.
(b) Admissible boundary conditions cannot specify separately the force
component at both ends of the same characteristic. If these forces
are unbalanced (after allowing for body forces), static equilibrium is lost.
\label{fig:pathfig}}
\end{figure}

Note that in
principle, one could have propagation also along the `backward'
characteristics (see Fig.~\ref{fig:pathfig}~(a)). This is forbidden since
these cut the free surface; any such propagation can only arise in the
presence of a nonzero surface force, in violation of the boundary
conditions. Therefore the fact that the propagation occurs only along
downward characteristics is not related to the fact that gravity acts
downward; it arises because we know already the forces acting at the free
surface (they are zero). Suppose we had instead an inverse problem:
a pile on a bed with some unspecified overload at the top surface, for
which the forces acting at the base had been measured. In this case, the
information from the known forces could be propagated along the {\em
upward} characteristics to find the unknown overload.
More generally, in {\sc osl} models, each characteristic ray
will cut the surface of a (convex) patch of material at two points; the sum of
the forces along the ray at the two ends must then be balanced by the
longitudinal component of the body force integrated along the ray 
(see Fig.~\ref{fig:pathfig}~(b)). These models are thus ``boundary fragile".

In three dimensions, the mathematical structure of these models is
somewhat altered,\cite{bcc} but the
conclusions are basically unaffected. The propagation of stresses is governed
by a Green's function which is the response to a localized overload; {\sc osl}
models predict that for (say) sand in a horizontal bed, the maximum
response at the base is not directly beneath a localized
overload but on a ring of finite radius (proportional to the depth) with this
as its axis.\cite{bcc}
(This could be difficult to test cleanly because of noise effects,
but there are related consequences for stress-stress correlations which are
discussed in Ref.~\cite{cargese,PRE}.) On the other hand, for different
geometries, such as sand in a bin, the stress propagation problem is not
well-posed even with hyperbolic equations, unless something is known about the
interaction between the medium and the sidewalls. But by assuming a
constant ratio to shear and normal forces at the walls, further interesting
predictions can be made, for example that the total weight increment
measured at the base of a cylindrical silo, in response to an overload on the
top, is a nonmonotonic function of the height of the fill.\cite{Claudin:Phd.thesis}
These predictions represent clear signatures of
hyperbolic stress propagation and, if confirmed experimentally, would be hard
to explain by other means.

\subsubsection{The problem of elastic indeterminacy}\label{subsec:Indet}

The well-posedness of the standard sandpile is not shared be models involving
the elliptic equations for an elastic body. For such a material, the stresses
throughout the body can be solved only if, at all points on the boundary,
either the force distribution or a displacement field is specified.\cite{Landau}
Accordingly, once the zero-stress boundary condition is applied at the
free surface, nothing
can in principle be calculated unless either the forces or
the displacements at the base are known (and the former amounts
to specifying in advance the solution of the problem).
The problem does not arise from any uncertainty about what to do
mathematically: one should specify a
displacement field at the base. Difficulties nonetheless arise if, as we
argued above, no physical significance can be attributed to this
displacement field for cohesionless poured sand.

\begin{figure}[hb]
\centerline{
\epsfig{file=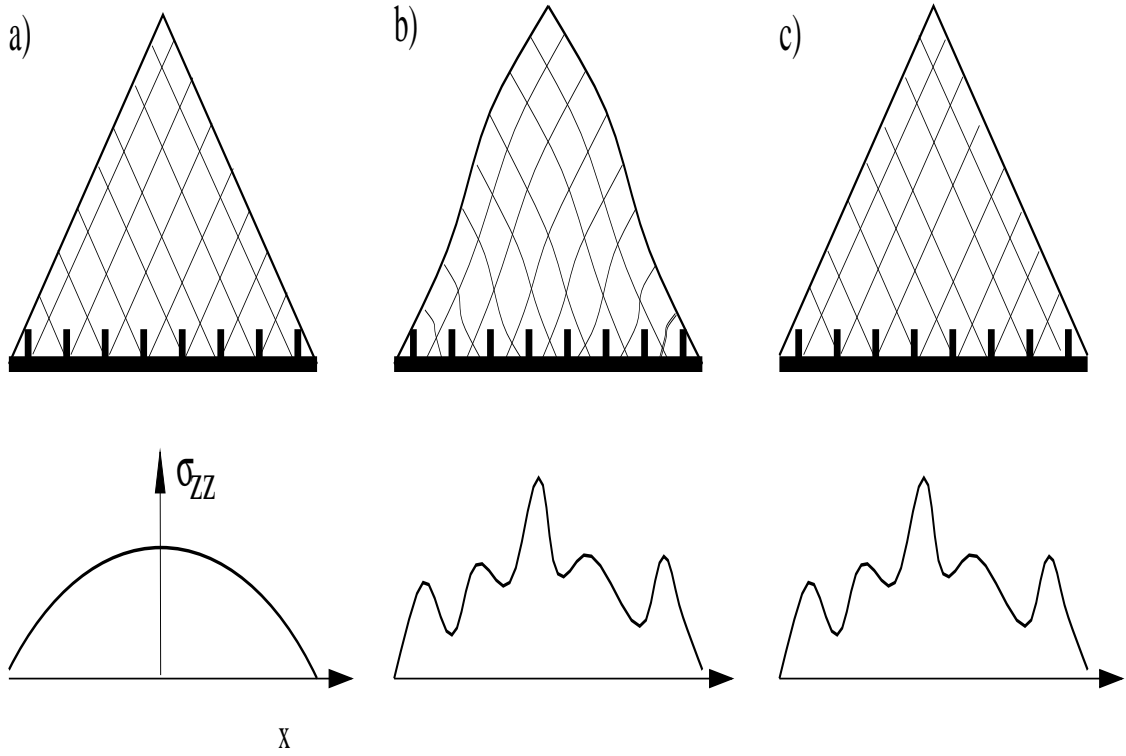,width=150mm,height=100mm,angle=0}}
\vspace*{2cm}
\caption[]{
Starting from an elastic cone or wedge on a rough
support, any initial stress distribution can be converted to
another by displacements with respect to the rough `pinning'
surface (a) $\to$ (b). Taking the limit of a high modulus (b)
$\to$ (c) at fixed surface forces, an arbitrary stress field remains,
while recovering the initial shape of the cone and satisfying the free
surface boundary conditions. This shows the physical character of
`elastic indeterminacy' for an elastic or elastoplastic body on
a rough support.
\label{fig:indeterminacy}}
\end{figure}

To give a specific example, consider the static equilibrium 
of an elastic cone of finite modulus, which is placed in an 
unstressed state (without gravity) onto a completely rough,
rigid surface; gravity is then switched on. This generates a pressure 
distribution with a smooth parabolic hump as in Fig.~\ref{fig:indeterminacy}a.
(The roughness can crudely be represented by a set of pins.)  
Starting from any initial configuration, another can be generated
by pulling and pushing parts of the body horizontally across the
base ({\em i.e.}, changing the displacements there); if this is rough,
the new state will still be pinned and will achieve a new static
equilibrium. This will generate a stress distribution, across the
supporting surface and within the pile, that differs from the
original one. Indeed, if a large enough modulus is now taken
(at fixed forces), this procedure allows one to generate arbitrary
differences in the stress distribution while generating
neither appreciable distortions in the shape of the cone, nor any
forces at its free surface.
This corresponds to a limit $Y \rightarrow \infty$,
$\underline{u} \rightarrow 0$ at fixed $Y \underline{u}$ where $Y$ is the 
modulus and $\underline{u}$ the displacement field at the base.

Analogous remarks apply to any simple
elastoplastic theory of sandpiles, in which an elastic zone, in
contact with part of the base, is attached at matching surfaces
to a plastic zone.  A natural presumption for the standard
sandpile might be that $Y\underline{u} = 0$ (that
is, the basal displacements vanish before the high modulus limit is taken).
This is consistent with the ``glued pile" interpretation of elastic models --
one assumes that glue also firmly attaches grains to the support as they
arrive. However, the same interpretation, as shown above, also requires
explicit consideration of quenched stresses (see Fig.~\ref{fig:floating}).
Note in any case that elastic and elastoplastic predictions for the sandpile
are indeterminate, in a rigorous mathematical sense, if the
$Y\to\infty$ limit is taken before the basal displacements $\underline{u}$
have been specified.

Experiments (reviewed in detail in Cates {\em et al.}\cite{RS})
report that for sandpiles on a rough rigid support, the forces on the base can
be measured reproducibly; and, although subject to statistical fluctuations on
the scale of several grains, do not vary too much
among piles constructed in the same way. 
In contrast, for any simple elastic or elastoplastic
model that does not include a specification of the basal displacements,
there is a very large indeterminacy in the predicted stress distribution,
even after averaging over any statistical fluctuations. 
An elastoplastic modeller who believes that the experiments measure something 
well-defined is then obliged to explain why and how the basal displacements 
(even if infinitesimal) are fixed by the construction history. 
Note that basal sag\cite{Savage98,Savage97}
is {\em not} a candidate for the missing mechanism, since it does
not resolve the elastic indeterminacy in these models; the latter arises
primarily from the {\em roughness}, rather than the rigidity, of the support.
An alternative view is that of Evesque,\cite{Evesqueprivate} who directly
confronts the issue of elastic indeterminacy and seemingly concludes that the
experimental results themselves {\em are and must be indeterminate}; he argues
that the external forces acting on the base of a pile are effectively chosen
at will, rather than actually measured, by the experimentalist 
(see also Ref.~\cite{EvesqueBoufellouh}). 
To what extent this viewpoint is based on experiment,
and to what extent on an implicit presumption in favour of elastoplastic
theory, is to us unclear.

\subsection{Crossover from Fragile to Elastic Regimes}\label{subsec:Anisotropic}

We have emphasized above the very different modelling assumptions of the
fragile and elast(oplast)ic approaches to granular media. However, we have
recently shown that hyperbolic fragile behaviour can be recovered from an
elastoplastic description by taking a strongly anisotropic limit.\cite{PRL}
Moreover, the crossover between elastic and hyperbolic
behaviour, at least for one simple model of the granular skeleton,\cite{PRL}
is controlled by the {\em deformability} of the granular particles.
For simplicity in this section, we restrict attention to the {\sc fpa} model.

The {\sc fpa} model describes, by definition, a material in which the shear
stress must vanish across a pair of orthogonal planes fixed in the
medium -- those normal to the (fixed)
principal axes of the stress tensor. According to the Coulomb
inequality (which we also assume) the shear stress must also be
less than
$\tan\phi$ times the normal stress, across planes oriented in all other
directions.  Clearly this combination of requirements can be viewed as a
limiting case of an elastoplastic model with an anisotropic yield
condition:
\be
|\sigma_{tn}| \le \sigma_{nn}\,\tan\Phi(\theta) \label{eq:yieldcrit}
\ee
where $\theta$ is the angle between the plane normal ${\bf n}$ and the
vertical (say) and ${\bf t \cdot n} = 0$.
An anisotropic yield condition should arise, in principle, in any
material having a nontrivial fabric, arising from its construction
history. The limiting choice corresponding to the {\sc fpa} model for a
sandpile is
$\Phi(\theta) = 0$  for $\theta = (\pi - 2\phi)/4$ (this corresponds to
planes where $\bf n$ lies parallel to the major principal axis), and
$\Phi(\theta) = \phi$ otherwise.
(There is no separate need to specify the second,
orthogonal plane across which shear stresses vanish, since this is
assured by the symmetry of the stress tensor.)
By a similar argument, all other {\sc osl} models can also be
cast in terms of an anisotropic yield condition, of the form $|\sigma_{tn} -
\sigma_{nn}\,\tan\Psi(\theta)| \le \sigma_{nn}\tan\Phi(\theta)$ where
$\Phi(\theta)$ vanishes,
and $\Phi(\theta)$ is finite for two values of $\theta$.
(This fixes a {\em nonzero} ratio of shear and normal
stresses across certain special planes.)

At this purely phenomenological level there is no difficulty in connecting
hyperbolic models smoothly onto {\em highly anisotropic} elastoplastic
descriptions.
Specifically, consider a medium having an orientation-dependent
friction angle $\Phi(\theta)$ that does not actually vanish,
but is instead very small
($\le \epsilon$, say) in a narrow range of angles (say of order
$\epsilon$) around $\theta = (\pi - 2\phi)/4$, and approaches $\phi$
elsewhere.  (One interesting way to achieve the required yield anisotropy is
to have a strong anisotropy in the {\em elastic} response, and then impose a
uniform yield condition to the strains, rather than stresses.)

Such a material will have, in principle, mixed
elliptic/hyperbolic equations of the usual elastoplastic type. The
resulting elastic and plastic regions must nonetheless arrange
themselves so as to obey the {\sc fpa} model to within terms that vanish as
$\epsilon \to 0$. If
$\epsilon$ is small but finite, then for this elastoplastic
model the results  will depend on the basal boundary condition,
but only through these higher order corrections to the leading ({\sc fpa})
result. Thus, although elastoplastic models do suffer
from elastic indeterminacy (they require a basal displacement field to be
specified), the extent of the influence of the boundary condition on the
solution depends on the model chosen. Strong enough
(fabric-dependent) anisotropy, in an elastoplastic description, might so
constrain the solution that it is {\em primarily the granular fabric} (hence
the construction history) and only minimally the boundary conditions which
actually determine the stresses in the body. For models such as that given
above there is a well-defined limit where the indeterminacy is entirely lifted,
hyperbolic equations are recovered, and it is quite proper to talk of
local stress propagation `rules' determined by the
construction history of the material.
Our continuum modelling framework is based precisely on these assumptions.

The crossover just outlined can also be understood directly in terms of the
micromechanics of force chains, at least within the simplified picture
developed in Section \ref{sec:FM}. We consider a regular lattice of force
chains (see Fig.~\ref{fig:jammed}~(b)), for simplicity rectangular (the {\sc fpa} 
case), which is fragile if the chains can support only longitudinal forces. 
As mentioned in Section \ref{subsec:Anisotropic}, 
this is true so long as such paths  consist of
linear chains of rigid particles, meeting at frictional point contacts: the
forces on all particles within each chain must then be colinear, to avoid
torques. This imposes the ({\sc fpa}) requirement that there are no shear
forces across a pair of orthogonal planes normal to the force chains
themselves. 
Suppose now a small degree of particle deformability is introduced.\cite{PRL}
This relaxes {\em slightly} the collinearity requirement, but only
because the point contacts are now flattened (see Fig.~\ref{fig:paths}~(b)). 
The ratio $\epsilon$ of the maximum transverse load to the normal one will
therefore vanish with (some power of) the mean compression. This
yield criterion applies only across two special planes; failure
across others is governed by some smooth yield requirement (such as
the ordinary Coulomb condition: the ratio of the principal stresses
lies between given limits). The granular skeleton just described, which
was fragile in the limit of rigid grains, is now governed by a strongly
anisotropic elastoplastic yield criterion of precisely the kind described above.
This indicates how a packing of frictional, deformable
rough particles, displaying broadly conventional elastoplastic features
when the deformability is significant, can approach a fragile limit
when the limit of a large modulus is taken. (It does
not prove that {\em all} packings become fragile in this
limit.) Conversely it shows how a packing that is basically fragile in
its response to a graviational load could nonetheless support very small
incremental deformations, such as sound waves, by an elastic mechanism.

The question of whether sandpiles are
better described as fragile, or as ordinarily elastoplastic, remains open
experimentally. To some extent it may depend on the question being
asked. However, we have argued, on various grounds, that in calculating the
stresses in a pile under gravity a fragile description may lie closer to the
true physics.

\section{Conclusions}\label{sec:Conclusion}

The jammed state of colloids, if it indeed exists in the laboratory, has not
yet been fully elucidated by experiment. It is interesting that even very
simple models, such as Eq.~(\ref{eq:nn}), can lead to nontrivial and testable
predictions (such as the constancy of certain measured stress ratios). Such
models suggest an appealing conceptual link between jamming, force chains, and
fragile matter.\cite{RS,PRL} However, further experiments are needed
to establish the degree to which they are useful in describing real
colloids.

For granular media, the existence of tenuous force-chain skeletons
is clear;\cite{Dantu,Thornton,Radjai,Kolymbas,Coppersmith}
the question is whether such skeletons are fragile. Several theoretical arguments
have been given, above and elsewhere, to suggest that this may be the
case, at least in the limit of rigid particles. Moreover, simulations
show strong rearrangement under small changes of compression axis; the
skeleton is indeed ``self-organized".\cite{Thornton,Radjai} 
Experiments also suggest cascades of  rearrangement \cite{staticavalanche,samchains}
in response to small disturbances. These findings are consistent with
the fragile picture.

The standard sandpile (a conical pile formed by pouring onto a rough rigid
support) has played a central role in our discussions. From the perspective of
geotechnical engineering, the problem of calculating stresses in the
humble sandpile may appear to be of only of marginal importance.
The physicist's view is different: the sandpile is
important, because it is one of the simplest problems in granular
mechanics imaginable. It therefore provides a test-bed for
existing models and, if these show shortcomings, may suggest
ideas for improved physical theories of granular media.

Given the present state of the data, a conventional elastoplastic
interpretation of the experimental results for sandpiles may remain tenable;
more experiments are urgently required.~\cite{Vanel99} 
In the mean time, a desire to keep using tried-and-tested modelling strategies 
until these are demonstrably proven ineffective is quite understandable. 
We find it harder to accept the suggestion\cite{Savage97}
that anyone who questions the general validity of traditional elastoplastic
thinking is somehow uneducated.

In summary, we have discussed a new class
of models for stress propagation in granular matter. These models assume
local propagation rules for stresses which depend on the construction
history of the material and which lead to hyperbolic differential
equations for the stresses. As such, their physical basis is
substantially different from that of conventional
elastoplastic theory.
Our approach predicts `fragile' behaviour, in which stresses
are supported by a granular skeleton of force chains that respond by
finite internal rearrangement to certain types of infinitesimal load.
Obviously, such models of granular matter might be incomplete in various ways.
Specifically we have
discussed a possible crossover to elastic behaviour at very small
incremental loads, and to conventional elastoplasticity at very high mean
stresses (when significant particle deformations arise). However, we believe
that our approach, by capturing at the continuum level at least some of the
physics of force chains, may offer important insights that lie beyond the
scope of previous continuum modelling strategies.

\section{Acknowledgment}

We thank 
R.~C.~Ball, E.~Clement, 
C.~S. and M.~J.~Cowperthwaite, 
S.~F.~Edwards, 
P.~Evesque,
P.-G. de Gennes, G.~Gudehus, J.~Goddard, 
D.~Levine, C.~E.~Lester,
J.~Melrose, S.~Nagel, 
F.~Radjai, J.-N.~Roux,
J.~Socolar, C.~Thornton, L.~Vanel and T.~A. Witten 
for illuminating discussions. 
Work funded in part by EPSRC (UK) GR/K56223 and GR/K76733.



\newpage

\newpage

\newpage

\newpage

\newpage

\newpage

\newpage

\newpage

\newpage

\end{document}